\documentclass{emulateapj}
\begin{document}
\title{CN Variations in High Metallicity Globular and Open
  Clusters\footnote{Based in part on data obtained at the W.M. Keck
    Observatory, which is operated as a scientific partnership among
    the California Institute of Technology, the University of
    California and the National Aeronautics and Space
    Administration. The Observatory was made possible by the generous
    financial support of the W.M. Keck Foundation.}} 

\bigskip
\author{Sarah L. Martell}
\affil{Astronomisches Rechen-Institut}
\affil{Zentrum f\"{u}r Astronomie der Universit\"{a}t Heidelberg}
\affil{69120 Heidelberg, Germany}
\and
\author{Graeme H. Smith} 
\affil{University of California Observatories/Lick Observatory}
\affil{Department of Astronomy \& Astrophysics}
\affil{University of California, Santa Cruz}
\affil{Santa Cruz, CA 95064}
\email{martell@ari.uni-heidelberg.de, graeme@ucolick.org}
\bigskip
\begin{abstract} 
\noindent
We present a comparison of CN bandstrength variations in the
high-metallicity globular clusters NGC 6356 and NGC 6528 with those
measured in the old open clusters NGC 188, NCG 2158 and NGC
7789. Star-to-star abundance variations, of which CN differences are a
readily observable sign, are commonplace in moderate-metallicity halo
globular clusters but are unseen in the field or in open clusters. We
find that the open clusters have narrow, unimodal distributions of CN
bandstrength, as expected from the literature, while the globular
clusters have broad, bimodal distributions of CN bandstrength, similar
to moderate-metallicity halo globular clusters. This result has
interesting implications for the various mechanisms proposed to
explain the origin of globular cluster abundance inhomogeneities, and
suggests that the local environment at the epoch of cluster formation
plays a vital role in regulating intracluster enrichment processes.

\end{abstract}


\section{Introduction}
Intracluster abundance variations are well-known among globular
cluster stars on the red giant branch, subgiant branch, and main
sequence (see Gratton et al. 2004\nocite{GSC04} for a thorough
review). Within halo globular clusters, at fixed stellar luminosity,
some stars are observed to have a typical Population II abundance
pattern, while others are depleted in carbon, oxygen, and magnesium,
and enriched in nitrogen, sodium, and aluminum. This latter abundance
pattern is most easily observed as unusually strong CN absorption in
low-resolution spectra, whereby it was first discovered. Consequently
such stars are referred to as ``CN-strong,'' and the entire abundance
pattern from carbon through aluminum is implied by that name. The
existence of CN-strong stars in globular clusters is presently
interpreted as a sign of an early enrichment process whose details are
murky (e.g., Cannon et al. 1998\nocite{C98}, Bekki et
al. 2007\nocite{B07}). This process apparently occurs in all halo
globular clusters \citep{GSC04}, but not in open clusters (e.g.,
Norris \& Smith 1985\nocite{NS85}, Smith \& Norris 1984\nocite{SN84}),
the halo field \citep{GSCB00}, or in the dwarf spheroidal satellites
of the Milky Way \citep{S03}. 

There are several possible explanations for the existence of CN-strong
stars in globular clusters, including pollution of cluster-forming gas
by AGB star winds (e.g., Cottrell \& Da Costa 1981\nocite{CD81}), an
early generation of stars with a top-heavy mass function (e.g., Cannon
et al. 1998\nocite{C98}), or Wolf-Rayet stars (e.g., Smith
2006\nocite{S06}). There is also an emerging paradigm that
sufficiently massive globular clusters form their stars in two
generations (e.g., Carretta et al. 2008\nocite{CB08} and D'Ercole et
al. 2008\nocite{DVD08}). In this picture, cluster mass loss is driven
by supernovae, primarily affecting the first generation of stars and
accounting for the varying ratios of halo-like to enriched stars and
differing horizontal branch morphologies. All of the various models
have a few common characteristics: the process must happen quickly,
since there is a limited time from the first star formation in a
globular cluster until the $0.8$M$_{\odot}$ stars that are presently
red giants have finished forming. All models rely on intermediate- to
high-mass stars as a source for polluting material, both because of
their relatively short lifetimes and because the existence of
anticorrelated Mg and Al abundances implies high-temperature
fusion. All models also require a proto-cluster to have a significant
gravitational potential (to retain enriched gas) and a sufficiently
rarefied environment (to prevent ram pressure stripping). Given these
characteristics, it can be understood qualitatively why old, isolated
halo globular clusters would contain CN-strong stars while lower-mass
disk objects like open clusters would not. 
 
Disk globular clusters, with their relatively high metallicities,
provide an interesting intersection of those two populations, and a
convenient testbed for models of the primordial enrichment process. If
the process is prohibited by in-disk orbits or high metallicity, then
disk globular clusters will have homogeneous abundance patterns
similar to those in open clusters. However, if the high initial mass
and relatively deep potential well of a proto-globular cluster is the
key to allowing enrichment, disk and bulge globular clusters could
also contain a subpopulation of stars with the same CN-strong
abundance pattern observed in halo globular clusters (if the yields of
the enriching stars scale with metallicity). 

It is in this context that we consider our data set, which is
comprised of Keck/LRIS blue spectra of 24 upper red giant branch stars
in the high-metallicity disk globular cluster NGC 6356 and 11 similar stars 
in NGC 6528. NGC 6356 is in a location consistent with being a disk
globular cluster (e.g., Zinn 1985\nocite{Z85}): 2.7 kpc above the
Galactic plane, on the far side of the Galactic center \citep{BOB94}. Its
metallicity is [Fe/H]$=-0.50$ according to the 2003 revision of the
\citet{H96} catalog. NGC 6528 is located in Baade's Window, 0.6 kpc
from the Galactic center and 0.6 kpc below the plane. As such, it is
considered a bulge globular cluster \citep{M95b}, with \citet{Z04} and
\citet{O05} finding metallicities of [Fe/H]$=-0.1$ and $-0.2$ dex from
optical and infrared spectroscopy, respectively. We compare these
spectra to Lick/Kast blue spectra of 22 red giants in the old open
clusters NGC 188 ([Fe/H]$=-0.06$, Carraro \& Chiosi
1994\nocite{CC94}), NGC 2158 ([Fe/H]$=-0.60$, Carraro et
al. 2002\nocite{CGM02}), and NGC 7789 ([Fe/H]$=-0.04$, Tautvai{\v
  s}ien{\.e} et al. 2005\nocite{T05}). These are all northerly
clusters, with estimated ages ranging from 1.3 Gyr for NGC 7789 to 7.5
Gyr for NGC 188 (both age estimates from Carraro \& Chiosi
1994\nocite{CC94}). We measure the strength of the CN absorption band
at $4215\mbox{\AA}$ with the index  $S(4142)$ defined by
\citet{NF79}. This quantity, like most spectral indices, measures the
magnitude difference between the integrated flux inside the absorption
feature in question and in a nearby continuum region. By considering CN
bandstrength variation as a proxy for the full enriched abundance
pattern, we hope to explore whether NGC 6356 and NGC 6528 exhibit CN
bimodality like their low-metallicity halo counterparts do, or whether
their similarity to open clusters includes a homogeneous abundance
distribution along with near-Solar metallicities and disk orbits.

\section{The Data Set}
Spectra in NGC 6356 were acquired on two nights in 2002 June, using
the Low Resolution Imaging Spectrometer (LRIS, Oke et
al. 1995\nocite{O95}) at the Keck I Telescope on Mauna Kea. Spectra in
NGC 6528 were taken on two nights in 2003 June with the same
instrumental setup. Using a mirror in place of a dichroic, all light
was directed to the mosaic of two 2Kx4K Marconi CCDs in the blue
camera. The 400/3400 grism and a slit width of $1\arcsec$
($8.7\arcsec$ for flux standard stars) resulted in a pixel spacing of
$1.1\mbox{\AA}$ and a resolution of $8\mbox{\AA}$. Typical exposure
times of 1200-1500 seconds produced a signal-to-noise ratio (S/N) per pixel
of roughly 60 at $4320\mbox{\AA}$, just redward of the G band. 

\begin{deluxetable}{l l r c c c}
\tablewidth{0pt}
\tablecaption{\label{HIZt1}}
\tablehead{\colhead{Cluster ID} &
\colhead{Star ID} &
\colhead{$M_{V}$} &
\colhead{B-V} &
\colhead{t(s)} &
\colhead{$S(4142)$}}
\startdata

NGC 6356\tablenotemark{a}   & 14         &   -0.81 &   1.51 &    2400 &  -0.119\\
           & 17         &   -1.33 &   2.11 &    1800 &  -0.123\\
           & 18         &   -0.59 &   1.42 &    1200 &  0.019\\
           & 20         &   -1.15 &   1.63 &    1200 &  0.078\\
           & 21         &   -0.93 &   1.42 &    1200 &  -0.034\\
           & 61         &   -0.89 &   1.69 &    1200 &  0.081\\
           & 69         &   -0.42 &   1.42 &    1500 &  -0.140\\
           & 70         &   -0.85 &   1.62 &    1200 &  -0.102\\
           & 71         &   -0.67 &   1.52 &    1200 &  0.052\\
           & 73         &   -1.18 &    1.7 &    1500 &  -0.099\\
           & 74         &   -0.43 &   1.55 &    1500 &  -0.136\\
           & 77         &   -0.87 &    1.7 &    1500 &  0.127\\
           & 93         &   -0.75 &   1.49 &    1500 &  -0.191\\
           & 9          &   -0.65 &   1.52 &    1200 &  0.141\\
           & 104        &   -0.45 &   1.65 &    1200 &  -0.149\\
           & 107        &    0.07 &   1.39 &    1500 &  0.081\\
           & 111        &    -1.1 &   1.76 &    1500 &  -0.053\\
           & 113        &    -1.0 &   1.66 &    1200 &  0.082\\
           & 116        &   -0.01 &   1.26 &    1500 &  0.002\\
           & 154        &   -0.55 &   1.64 &    1200 &  -0.148\\
           & 157        &   -0.89 &   1.49 &    1200 &  -0.157\\
           & 164        &   -0.68 &   1.71 &    1500 &  -0.052\\
           & 166        &   -0.55 &   1.69 &    1200 &  -0.115\\
           & 53         &   -0.44 &   1.36 &    1500 &  0.119\\
           & 81         &   -0.24 &   1.46 &    1500 &  0.081\\
NGC 6528\tablenotemark{b}   & 0-16       &  -0.031 &    1.57    &    2100 &  -0.067\\
           & II-19       &   0.027 &   1.73     &    2100 &  -0.027\\
           & II-22       &  -0.012 &   1.70     &    2100 &  0.118\\
           & II-51       &   0.125 &   1.78     &    1800 &  -0.069\\
           & II-7        &  -0.409 &    1.88    &    3000 &  -0.037\\
           & II-8        &  -0.345 &    1.89    &    1800 &  0.083\\
           & II-12       &  -0.279 &   1.65     &    1800 &  -0.094\\
           & II-27       &   0.206 &   1.66     &    2100 &  0.027\\
           & II-39       &   -0.15 &    1.85    &    1800 &  0.083\\
           & II-3        &  -0.157 &    1.74    &    1800 &  -0.059\\
           & II-4        &   0.101 &    1.53    &    2100 &  -0.052\\
           & II-70       &  -0.179 &   1.88     &    3900 &  -0.045\\
\enddata
\tablenotetext{a}{NGC 6356: Star identifiers and photometry are taken
  from \citet{SW60}.} 
\tablenotetext{b}{NGC 6528: Star identifiers and photometry are from
  \citet{VY79}.} 
\end{deluxetable}

The open cluster stars were observed with the Shane 3m telescope at
Lick Observatory on two nights in 2005 February, three nights in 2005 September, and three nights in 2006 August. In each case a mirror
was used to direct all light to the blue side of the Kast double
spectrograph. The 600/4310 grism and a slit width between $1\arcsec$
and $2\arcsec$, depending on weather ($9\arcsec$ for flux standard
stars), produced a pixel spacing of 1.8$\mbox{\AA}$ and a resolution
of 5.4$\mbox{\AA}$. Exposure times varied widely because of a range of
two magnitudes in distance modulus, and the typical signal-to-noise
ratio at $4320\mbox{\AA}$ is 150. Table \ref{HIZt1} lists identifying
information, photometry, exposure time and the CN bandstrength index
$S(4142)$ for each globular cluster giant observed. Analogous
information for the open cluster giants, and membership probabilities
where available, are given in Table \ref{HIZt2}. 

\begin{deluxetable*}{l l r c c c c c}
%\tablewidth{0pt}
\tablecaption{\label{HIZt2}}
\tablehead{\colhead{Cluster ID} &
\colhead{Star ID} &
\colhead{$M_{V}$} &
\colhead{B-V} &
\colhead{t(s)} &
\colhead{S(4142)} &
\colhead{$\rm{P_{pm}}$ (\%)} &
\colhead{$\rm{P_{rv}}$ (\%)}}
\startdata

NGC 188\tablenotemark{a}    & II-76      &    0.64 &   1.23 &    2700 &  0.145 &     84 &    98\\
           & I-69       &    0.47 &   1.34 &    2700 &  0.020 &     86 &    96\\
           & I-105      &    0.56 &   1.25 &    2700 &  0.052 &     85 &    98\\
           & II-72      &    0.62 &   1.33 &    2700 &  0.074 &     85 &    98\\
           & III-18     &   -0.44 &   1.51 &    1080 &  -0.054 &     82 &    98\\
           & I-1        &    0.12 &   1.18 &    2400 &  0.031 &     82 &    98\\
           & II-51      &    1.14 &   1.18 &    2700 &  -0.037 &     81 &    98\\
NGC 2158\tablenotemark{b}   & S1 R1 16   &   -1.43 &   1.85 &    5400 &  0.015 &  ...   &  ... \\
           & S3 R1 32   &   -1.72 &   1.79 &    5400 &  -0.037 &   ...  & ...  \\
           & S4 R1 8    &   -1.54 &   1.63 &    5400 &  -0.084 &   ...  & ...  \\
           & S2 R4 c    &   -1.47 &   1.62 &    7200 &  -0.035 &   ...  & ...  \\
           & S3 R2 55   &   -1.51 &   1.73 &    5400 &  0.018 &  ...   & ...  \\
           & S4 R5 12   &   -1.01 &   1.74 &    8100 &  -0.056 &   ...  & ...  \\
NGC 7789\tablenotemark{c}   & 491        &   -0.27 &   1.47 &    2400 &  -0.046 &  ...   &    97\\
           & 589        &   -1.42 &   1.65 &    1500 &  -0.009 &  ...   &    97\\
           & 614        &    -1.3 &   1.71 &    3600 &  -0.063 &  ...   &    97\\
           & 732        &   -1.06 &   1.63 &    2400 &  -0.004 &  ...   &    98\\
           & 583        &   -1.25 &   1.66 &     720 &  -0.006 &  ...   &    97\\
           & 654        &   -0.23 &   1.44 &     720 &  0.012 &  ...   &    84\\
           & 671        &   -0.45 &   1.48 &    1800 &  -0.005 &  ...   &    98\\
           & 933        &    -0.7 &   1.55 &    1200 &  -0.002 &  ...   &    98\\
           & 990        &    -0.6 &   1.51 &    1200 &  -0.009 &  ...   &    98\\
           & 1074       &   -0.67 &   1.56 &    1200 &  -0.011 &  ...   &    98\\
\enddata
\tablenotetext{a}{NGC 188: Star identifiers and photometry are taken
  from \citet{S62}, proper motion-based membership probabilities are
  from \citet{SMV04}, radial velocity-based membership probabilities
  are from \citet{G08}. } 
\tablenotetext{b}{NGC 2158: star identifiers and photometry are taken
  from \citet{AC62}. } 
\tablenotetext{c}{NGC 7789: Star identifiers and photometry are taken
  from \citet{MS81}, radial velocity-based membership probabilities
  are from \citet{G98}.} 
\end{deluxetable*}

\begin{figure} 
\begin{center}
\includegraphics[width=9.0cm]{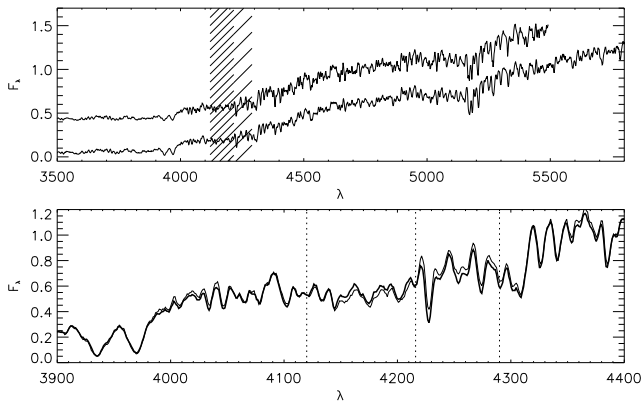}
%\plotone{f1.eps}
\end{center} 
%[Representative spectra from the data set]
\caption{
\textit{Upper panel}: representative spectra from the data set. The upper spectrum is NGC
7789 star 614, taken with the Kast spectrograph at Lick
Observatory. The lower spectrum is NGC 6356 111, taken with the LRIS
spectrograph at Keck Observatory. The closely spaced shaded region
shows the science band of the index $S(4142)$, and the more widely
spaced shading marks the $S(4142)$ comparison band. \textit{Lower panel}: comparison of CN-strong and CN-eak spectra in the region of the $4215 \mbox{\AA}$ CN band. The thinner line is the spectrum of NGC 6356 113, which has a stronger CN band than the heavier line, which is the spectrum of NGC 6356 111.
\label{HIZf1}}
\end{figure}

Data reduction was accomplished for both the Keck and Lick data sets
using the XIDL ``low-redux'' programs\footnote{Produced by
  J.X. Prochaska and available at
  http://www.ucolick.org/\char126xavier/LowRedux/}. The
instrument-specific programs handle bias subtraction, flat-fielding,
cosmic ray removal, object identification and extraction, sky
subtraction, flexure correction, wavelength calibration, atmospheric
correction, coadding, and flux calibration. The standard stars Feige
34, Feige 110, BD +28 4211, and BD +33 2642 were used for flux
calibration, with ``true'' spectra taken from the CALSPEC database
\citep{B96}. Figure \ref{HIZf1} shows a typical flux-calibrated
spectrum from each of the Kast (open cluster) and LRIS (globular
cluster) data sets, with the science band ($4120\mbox{AA}$ to
$4216\mbox{AA}$) and comparison band ($4216\mbox{AA}$ to
$4290\mbox{AA}$) of $S(4142)$ shown as regions with closely spaced
shading and widely spaced shading respectively.

\subsection{Cluster Membership}
Because all of the star clusters observed for this study orbit within
the disk or bulge, the line of sight to each of them contains more
foreground stars than the line of sight to a typical halo globular
cluster. Because of the near-solar metallicities of the cluster stars
and the typically low systemic radial velocities of the clusters
relative to the Sun, two of the most straightforward methods for
identifying foreground stars in low-metallicity globular cluster
fields are ineffective with the present data set. Three of our five
target lists (for NGC 6356, NGC 6528, and NGC 2158) were constructed
from photographic studies without the astrometric or spectroscopic
information necessary to identify foreground late-K and M dwarfs.

We use a combination of visual spectrum inspection, photometry, and
spectral indices focused on gravity-sensitive features to identify
foreground dwarfs and other non-useful stars in our data set and
remove them from further analysis. We reject NGC 6356 17 because of
TiO bands not seen in any of the other NGC 6356 stars, and NGC 6356 51
because it is significantly bluer and has a distinctly smaller Ca K
index \citep{M03} than all other stars observed in NGC 6356. We also
drop NGC 6528 II-27 from the sample because its $\lambda 3883$ CN
band, measured by the index $S(3839)$ \citep{B90}, is abnormally
small. In NGC 188, though the stars are all confirmed as cluster
members by proper motion \citep{SMV04} and radial velocity
\citep{G08}, star 1001 is thought to be a post-coalescence FK Comae
star \citep{HM85} and star 3018 is significantly brighter than the
rest of the sample. We are therefore unsure whether 3018 is an AGB
star, and we remove it from the analysis. NGC 7789 also has proper
motion-based membership probabilities in the literature
\citep{G98}. We do not see any outliers in the photometry or measured
indices in NGC 7789, and include all of the NGC 7789 stars listed in
Table \ref{HIZt2} in the analysis. In all, we exclude 5 of the 62
stars initially observed from the final sample, leaving 24 stars in
NGC 6356, 11 in NGC 6528, 6 in NGC 188, 6 in NGC 2158, and 10 in NGC
7789.

\section{CN Bandstrength Distribution}
\begin{figure} 
\begin{center}
\includegraphics[width=9.0cm]{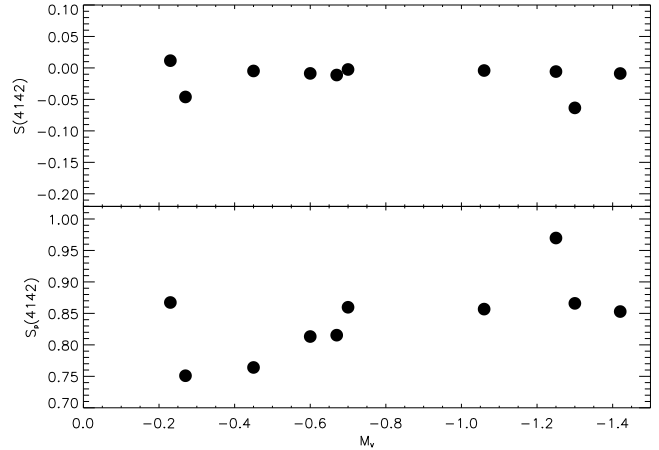}
%\plotone{f2.eps}
\end{center} 
%[CN index $S(4142)$ and pseudo-index $S_{p}(4142)$ in NGC 7789]
\caption{ 
Values of the CN index $S(4142)$ (\textit{upper panel}) and the pseudoindex
$S_{p}(4142)$ (\textit{lower panel}) measured for red giants in NGC 7789. As
described in the text, the RMS scatter in $S_{p}(4142)$ is used for
estimation of measurement error on the index $S(4142)$.  
\label{HIZf2}}
\end{figure}

We measure the CN bandstrength index $S(4142)$ from each of our
flux-calibrated spectra. Because it is defined as the magnitude
difference between the integrated flux inside the $4215\mbox{\AA}$ CN
band and a nearby CN-free region, larger values of $S(4142)$ reflect
stronger CN absorption. We determine the error on the index
measurement on a cluster-by-cluster basis. We define a pseudoindex
$S_{p}(4142)$, with the same comparison band as the original $S(4142)$
and the science band shifted to 4027\mbox{\AA} through 4077\mbox{\AA},
a CN-free region of the spectrum. The scatter in $S_{p}(4142)$ at a
given absolute magnitude will then be independent of CN bandstrength,
and will be representative of the scale of the random errors in
$S(4142)$. Figure \ref{HIZf2} shows $S(4142)$ (upper panel) and
$S_{p}(4142)$ (lower panel) versus $M_{V}$ for our NGC 7789 stars. As
expected from previous abundance-inhomogeneity studies of open
clusters, the stars stay within a fairly small range in $S(4142)$. In
NGC 7789 the scatter in $S_{p}(4142)$, which is $0.05$, is comparable
to that in $S(4142)$, implying that there is very little CN abundance
variation among the stars. It is also worth noting that $S_{p}(4142)$
rises with brightening $M_{V}$ in Figure 2, while $S(4142)$ does
not. This could be a result of the larger wavelength range spanned by
the bandpasses of $S_{p}(4142)$, which makes it more sensitive to
changes in stellar temperature. 

This pseudoindex error, calculated as the RMS of $S_{p}(4142)$ about
a best-fit line in the $S_{p}(4142)$ - $M_{V}$ plane, is calculated
for each of the clusters in our sample individually. For NGC 6356 the
scatter in $S_{p}(4142)$ is $0.048$ and for NGC 6528 the scatter is
$0.082$; a representative value of $\pm 0.05$ magnitudes is adopted as
the error in all measured values of $S(4142)$. After inspecting some
of the spectra, we feel that large-scale variations in continuum shape
are most likely to be responsible for the variations in
$S_{p}(4142)$. These may be a result of atmospheric diffraction
causing the scattering of blue light preferentially out of the slit,
particularly in NGC 6356 and NGC 6528, which are fairly southerly for
observation from the Keck I telescope. Since the bandpasses of
$S_{p}(4142)$ are more widely separated in wavelength than those of
$S(4142)$, and extend further into the violet,
the former index will be more sensitive to observational 
effects such as differential light loss at the spectrograph slit,
and errors in flux calibration. As such, our approach may overestimate 
the observational error in $S(4142)$ for the globular cluster 
observations.

\begin{figure} 
\begin{center}
\includegraphics[width=9.0cm]{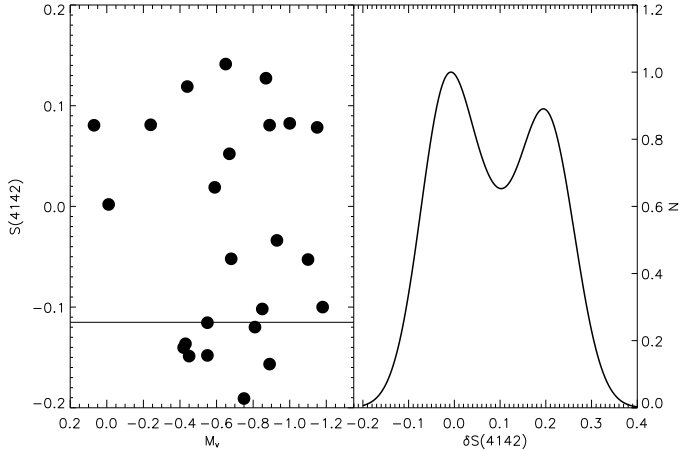}
%\plotone{f3.eps}
\end{center} 
%[CN bandstrength versus absolute V magnitude in NGC 6356]
\caption{
Measured values of the CN index $S(4142)$ in NGC 6356 (\textit{left panel}) and
a generalized histogram of the offset $\delta S(4142)$ between the
data points and the horizontal baseline (\textit{right panel}). 
\label{HIZf3}}
\end{figure}

\begin{figure} 
\begin{center}
\includegraphics[width=9.0cm]{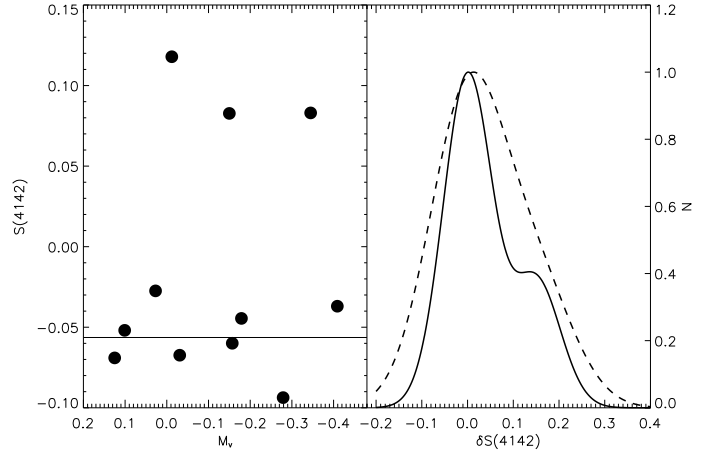}
%\plotone{f4.eps}
\end{center}
%[CN bandstrength versus absolute V magnitude in NGC 6528]
\caption{
Measured values of the CN index $S(4142)$ in NGC 6528 (\textit{left panel}) and
a generalized histogram of the offset $\delta S(4142)$ between the
data points and the horizontal baseline (\textit{right panel}). The presence of
a smaller gap between relatively CN-stronger and CN-weaker groups than in
NGC 6356, and the smaller number of CN-stronger stars, create a less
bimodal-looking generalized histogram than in Figure \ref{HIZf3}. The
dashed curve shows the generalized histogram calculated with a
$\sigma$ of $0.08$.
\label{HIZf4}}
\end{figure}

The left panel of Figure \ref{HIZf3} shows the CN bandstrength index
$S(4142)$ versus absolute visual magnitude for NGC 6356. There is
clearly a vertical range in $S(4142)$ at each $M_{V}$, which is also
seen in every other CN-bimodal globular cluster. Adapting the analysis
method of the \citet{N81} study of CN variations in NGC 6752, we fit a
line to the lower collection of points and measure the quantity
$\delta S(4142)$, the vertical distance from each point to the
line. In this study we set the slope of the baseline to zero, in
keeping with previous studies on 4215\mbox{\AA} CN bandstrength in
high-metallicity globular clusters (e.g., Briley
1997\nocite{B97}). The right panel of Figure \ref{HIZf3} shows a
generalized histogram of $\delta S(4142)$ values measured from the
left panel. This curve is produced by representing each $\delta
S(4142)$ point as a Gaussian with a width $\sigma$ equal to the
measurement error, which in this case is 0.05. The sum of those
individual Gaussians is then a realistically smoothed
histogram. Figure \ref{HIZf4} shows the analogous data in NGC
6528, with a generalized histogram calculated in the same way. With
fewer data points and a smaller gap in $S(4142)$ between 
relatively CN-weaker and CN-stronger groups, the generalized histogram
for NGC 6528 looks less bimodal than the generalized histogram for NGC
6356. The dashed curve in the
right panel of Figure \ref{HIZf4} shows the generalized histogram that
results if 
$\sigma$ is set to $0.08$, the measured standard deviation in
$S_{p}(4142)$ in NGC 6528. Increasing $\sigma$ smooths the curve
dramatically, reducing what appears in the solid curve to be a
distinct CN-stronger group to a broad, asymmetric distribution.

\begin{figure} 
\begin{center}
\includegraphics[height=10.0cm]{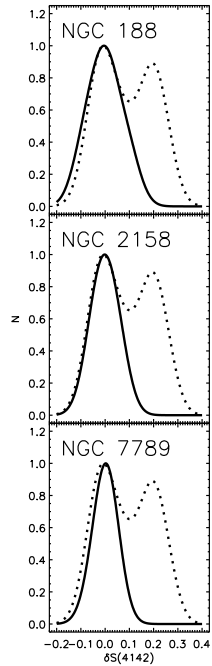}
%\plotone{f5.eps}
\end{center} 
%[Generalized histograms of $\delta S(4142)$ in all globular and open clusters in this study]
\caption{ 
Generalized histograms of the CN bandstrength index $\delta S(4142)$
for each open cluster in this survey, with the generalized histogram of
$\delta S(4142)$ in NGC 6356 overplotted as a dotted line. The
open clusters all have single-peaked distributions in CN bandstrength,
in clear contrast to the disk globular cluster NGC 6356. 
\label{HIZf5}}
\end{figure}

Figure \ref{HIZf5} shows generalized histograms of $\delta
S(4142)$ for the open clusters NGC 188, NGC 2158, and NGC 7789, all calculated
in the same way as for the globular clusters, with flat baselines and
$\sigma$ set 
to 0.05. The curves are all normalized at the peak for easier
intercomparability, and the generalized histogram for NGC 6356 is
overplotted in each panel as a dotted curve. It is
clear that the distribution of CN bandstrength in the three open 
clusters is single-peaked, and that there are two groups in CN
bandstrength in NGC 6356, but NGC 6528 is more difficult to
categorize. It is possible that the difference in [N/Fe]
between CN-enhanced and CN-normal stars is smaller in Solar-metallicity
globular clusters, since the total mass in nitrogen required to
produce an abundance gap of up to 1 dex is far larger than in
low-metallicity globular clusters. The broad but not clearly bimodal
distribution of CN bandstrengths in NGC 6528 could be a sign that
there is an upper limit on the amount of nitrogen that can be injected
into feedback material in high-metallicity globular clusters.

\section{Results}
The homogeneity of CN bandstrengths in the open clusters surveyed is
an expected result: \citet{NS85}, \citet{SN84}, and \citet{T05} have
all demonstrated small ranges in bandstrengths or abundances in these
clusters. However, the disk/bulge globular clusters NGC 6356 and NGC
6528 have distributions of CN bandstrength that clearly resemble those
observed in halo globular clusters like NGC 6752 and M3, and that is
surprising. We interpret the presence of CN-strong stars in NGC 6356
and NGC 6528 to mean that the same primordial enrichment process took
place in disk and bulge globular clusters as in halo globular
clusters. Higher resolution spectroscopy is needed to determine
whether the full ``enhanced'' abundance pattern from carbon through
aluminum is present in the CN-strong stars. On a related note, \citet{O08}
have found evidence of a Mg-Al abundance anticorrelation, such as
often accompanies CN variations in globular clusters, in the bulge
globular NGC 6441 ([Fe/H] $= -0.5$). 

The presence of CN-strong stars in disk and bulge globular clusters
provides strong constraints on the primordial enrichment
process. Since the Galactic orbits of NGC 6356 and NGC 6528 do not
allow for long periods of time outside the disk or bulge, the transfer
of enriched material from evolved higher-mass stars to still-forming
low-mass stars must have been both relatively fast and efficient in
order to avoid ram pressure stripping from a primordial environment
that was likely much denser than at present. In addition, the absolute
amount of nitrogen needed to create a difference of between $0.3$ and
$1.0$ dex in [N/Fe] between CN-weak and CN-strong stars is
significantly larger at near-Solar metallicity than it is at typical
halo metallicity. This implies that the material transferred to
low-mass stars must have been more enriched in nitrogen in the
higher-metallicity globular clusters. Such a requirement may pose a
problem for AGB star enrichment scenarios. The 4 and 5 $M_{\odot}$ AGB
models with Reimers mass loss rates of \citet{KL07} eject a total mass
of $^{14}$N of $7.7 \times 10^{-2}$ $M_{\odot}$ and $3.6 \times
10^{-2}$ $M_{\odot}$ respectively for $Z=0.0001$, $2.3 \times 10^{-2}$
$M_{\odot}$ and $6.3 \times 10^{-2}$ $M_{\odot}$ respectively for
$Z=0.004$, and $2.2 \times 10^{-3}$ $M_{\odot}$ and $5.4 \times
10^{-2}$ $M_{\odot}$ respectively for $Z=0.008$. There is no
indication from these models that early AGB stars in bulge or disk
globular clusters would eject a proportionally larger amount of
nitrogen than intermediate-mass AGB stars in halo clusters. To obtain
comparable enhancements of [N/Fe] in CN-strong giants in both disk and
halo globular clusters, it may be necessary to assume that the former
were enriched by a larger number of AGB stars per unit cluster mass
than halo clusters.  

The evidence in this article indicates that disk and bulge globular
clusters contain a population of CN-enhanced giants, presumably
analogous to those in halo globular clusters, whereas open clusters of
similar metallicity do not. This may indicate that open clusters were
not massive enough to experience early enrichment of the type evinced
by globular clusters. Alternatively, disk and bulge globular clusters
may have had a larger than typical number of AGB stars per unit
mass. Could this have been a consequence of the environment in which
the bulge and disk globulars formed? Conditions in the central parts
of the proto-Galaxy may have favored high- and intermediate-mass star
formation under starburst conditions. Open clusters, forming further
out in the Galactic disk in more quiescent molecular clouds, could
have resulted from star formation activity more conducive to
lower-mass stars. 

Indeed, this consideration of AGB-star yields suggests that
self-enrichment of disk and bulge globulars may been have associated
with more top-heavy stellar mass functions than were present even in
halo globular clusters. Many halo globular clusters may have formed
within proto-dwarf galaxies or massive clouds of the type suggested by
\citet{SZ78}, and their chemical evolution would have been set by the
environment within such systems. Recent studies of multiple stellar
populations in Galactic and extragalactic globular clusters, such as
\citet{M08}, \citet{P07}, and \citet{MBN08}, make it clear that their
star formation took place in a complex and dynamic environment. The sites of disk and bulge cluster formation may have been strongly influenced by conditions within the chaotic and violent inner regions of the proto-Galaxy. Cloud collisions may have been fairly common in the still-forming bulge, resulting in starbursts that were relatively efficient in high- and intermediate-mass star formation, leading to pronounced chemical enrichment that built up not only high overall metallicities, but also high [N-Na-Al/Fe] ratios as well. As such, the study of abundance
inhomogeneities in disk and bulge globular clusters could offer the
promise of probing some of the most efficient primordial environments
for chemical evolution.  

\acknowledgments{
The authors wish to recognize and acknowledge the very significant
cultural role and reverence that the summit of Mauna Kea has always
had within the indigenous Hawaiian community.  We are most fortunate
to have the opportunity to conduct observations from this mountain. 

This research has made use of the WEBDA database, operated at the
Institute for Astronomy of the University of Vienna.}

%\bibliography{abund_master}

\end{document}